\documentstyle[epsfig]{article}

\begin{document}
\centerline{PUTTING A SPIN ON THE}
\centerline{AHARONOV-BOHM OSCILLATIONS}
\bigskip
\centerline{Jeeva Anandan}
\smallskip

\centerline {Department of Physics and Astronomy }
\centerline {University
of South Carolina}
\centerline {Columbia, SC 29208, USA }

\centerline{and}

\centerline{Clarendon Laboratory}
\centerline{University of Oxford}
\centerline{Parks Road}
\centerline{Oxford, ~~UK}

\centerline{E-mail: jeeva@sc.edu}
\medskip
\centerline{May 30,2002, revised June 8, 2002.}
\bigskip\noindent
{\it A slightly different version was published in Science, Vol. 297, 1656  (Sept. 6, 2002).}
\bigskip

Of the many fascinating consequences of quantum mechanics, among
the more mysterious ones is the Aharonov-Bohm effect. According to
classical physics, a charged particle would be influenced by a
magnetic field only if the particle goes through a region in which
the magnetic field strength is nonzero. But according to quantum
mechanics, if the quantum wave representing the state of a charged
particle, such as an electron, is split into two waves that go
around a solenoid and interfere, the resulting interference
pattern is influenced by the magnetic flux enclosed by the waves,
even though the particle is nowhere in the region of non vanishing
field strength. This was predicted by Aharonov and Bohm in one of
the most influential papers of physics in the latter part of the
twentieth century \cite{ah1959}. In addition to its charge, the
electron has a magnetic moment proportional to its spin. So, if
the magnetic field strength is non vanishing along the electron
wave, then in addition to the above flux dependent effect, there
are also spin dependent effects due to the interaction of the
magnetic moment with the magnetic field. If in addition an
electric field is present, there is the so the called spin-orbit
interaction of the magnetic moment with the electric field
\cite{sa1994}, which further influences the interference pattern
\cite{zh2000}. All these effects are present in a recent beautiful
experiment performed by Yau, De Poortere and Shayegan
\cite{ya2002,mo1998}.

\centerline{\psfig{figure=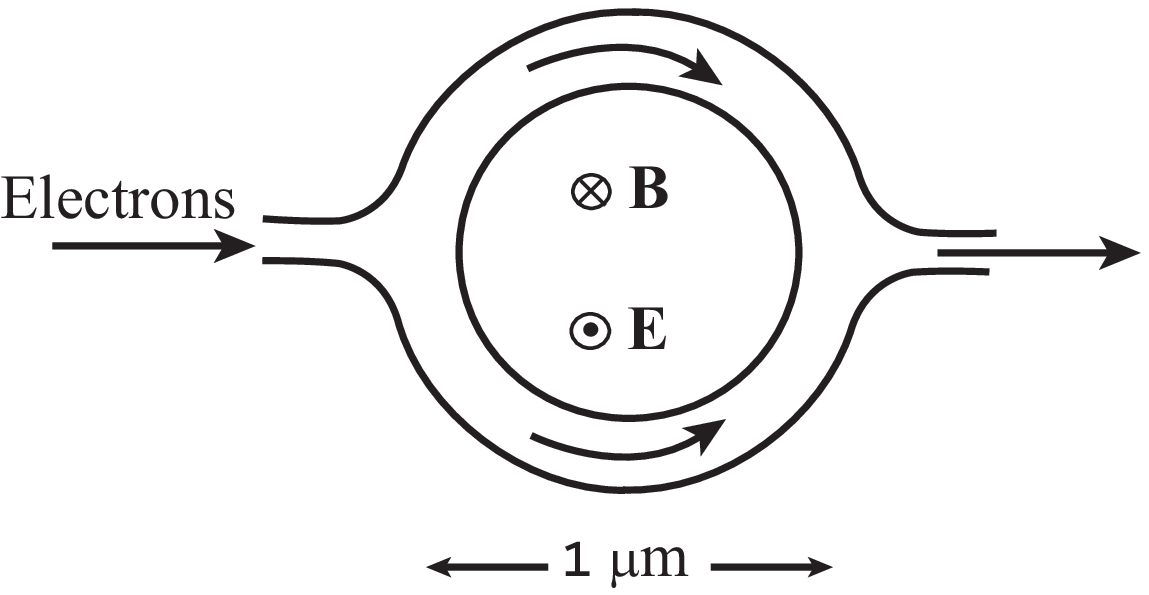,height=2in}}

\bigskip

\noindent Figure 1. {\bf Electrons enter and leave the ring
structure as indicated by the horizontal arrows.  The quantum wave
associated with each electron in the entrance region splits into
two wave packets that go around the ring, as indicated by the
curved arrows, and interfere in the exit region. Homogeneous
magnetic field $\bf B$ and electric field $\bf E$ are applied
normal to the plane of the ring. Whether the interference is
constructive or destructive, and correspondingly whether the
current is maximum or minimum, depends on the values of $\bf B$
and $\bf E$. Hence, varying these fields leads to oscillations of
the current.}

\bigskip

An interesting aspect of this experiment is that it is a
miniaturized version of the original experiments that confirmed
the Aharonov-Bohm effect \cite{pe1989}. The apparatus consists of
a ring structure (Fig. 1) with a diameter of only about one $\mu$m
(micro-metre or one- thousandth of a millimetre) that is
fabricated inside a GaAs/AlGaAs heterostructure.
This is an example of the new field of mesoscopic physics that
deals with structures that are intermediate between atomic and
macroscopic scales. The linear dimensions of the tiny apparatus
used in mesoscopic physics vary from about a $\mu$m to a
nano-metre, which is only about ten times the size of an atom. If
such a tiny apparatus made of metals or semi-conductors is made
sufficiently cold (about $30$ milli-kelvin)
then the conduction electron waves are coherent over the entire
apparatus, because there is no randomization due to inelastic
scattering. This leads to interesting quantum effects. In the
present experiment, the conduction electrons are like a quantum
gas that enters and exits the ring (Fig. 1), and therefore
constitutes a current. Homogeneous magnetic and electric fields
are applied normal to the plane of the ring structure. The
electric field $\bf E$ is needed to confine the electrons to this
plane.

It was predicted \cite{al1981} that the electrical resistance of
the current through a mesoscopic ring would vary as the magnetic
flux $\Phi$ through the ring is varied by changing the strength of
the magnetic field $\bf B$. Moreover, this magneto-resistance is
an oscillatory function of the magnetic flux with period $h/e$,
where $h$ is Planck's constant and $e$ is the charge of the
electron. This is a consequence of the fact that the Aharonov-Bohm
phase shift is ${2\pi e\over h}\Phi$, and the current is maximum
when there is constructive interference and minimum when there is
destructive interference, for a given externally applied
potential. This has been well confirmed by experiments
\cite{wa1986}, including the present one (Fig. 2). This means that
if one writes the magneto-resistance function $R(\Phi)$ as a sum
of all possible (normalized) periodic functions of $\Phi$, then
the coefficients of this expansion that multiply the functions
whose period $\tau$ is equal to or very close to $h/e$ are very
large compared to the other coefficients. In other words, the
Fourier transform of $R$ that is the function $\tilde R(f)$ of the
frequency $f=1/\tau$ (whose values are the coefficients of the
above expansion) would be highly peaked at $f=e/h$.

\centerline{\psfig{figure=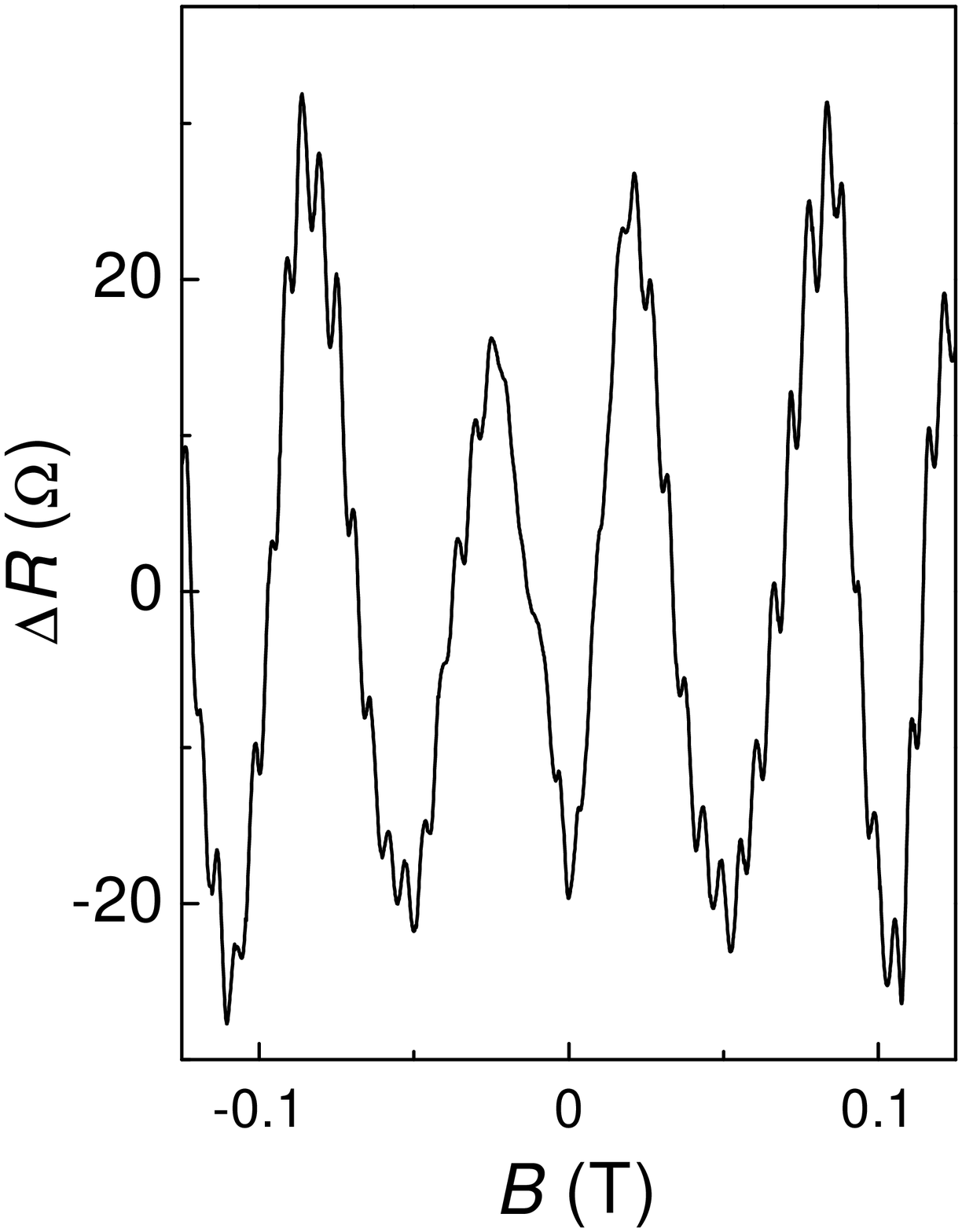,height=3.5in}}

\bigskip

\noindent Figure 2. {\bf $\Delta R$ is the resistance (in Ohms),
after subtracting away a smooth background resistance, in the
presence of an externally applied magnetic field $B$ (in Teslas).
As $B$ is varied (with the externally applied electric field
fixed), $\Delta R$ oscillates. This is due to the oscillation of
the current through the ring structure in Fig. 1 that is inversely
proportional to the resistance, for a constant applied potential
difference, according to Ohm's law.} (This graph is taken from
ref. 4.)

\bigskip

In the experiment of Yau et al \cite{ya2002}, this peak was
observed. But in addition two other smaller peaks on either side
of the main peak were also observed, suggesting a modulation of
the Aharonov-Bohm oscillation of $R$. It is reasonable to regard
these side peaks as being due to spin-dependent effects. These may
be obtained by supposing that the electron interacts with an
effective magnetic field \cite{sa1994} that is $ {\bf {\cal B}}=
{\bf B} -{1\over 2}{\bf v}\times {\bf E}$, where $\bf v$ is the
`velocity' of the electron in the approximation that its wave is
approximately a plane wave. The factor $1\over 2$ distinguishes
the electron from the corresponding effective magnetic field
experienced by a neutral dipole, such as the neutron, in the
combined magnetic and electric fields.
For an electron in an atom, this difference is traditionally
attributed to the semi-classical Thomas precession \cite{sa1994}.
It should be noted that the $\bf E$ experienced by the electron is
much larger than the applied electric field. This has been
determined to be the case experimentally, and has been attributed
to complex band effects in the semi-conductor \cite{by1984}.

The above combined effect of the electric and magnetic fields on
the electrons may also be regarded as arising from Berry's phase
\cite{be1984}. This is a geometric phase acquired by the wave
function when it evolves slowly (adiabatically) and returns to the
original state. In the present case, because $ {\bf {\cal B}}$
varies slowly during the motion of the electron along each
semi-circular ring, the components of the spin state of each
electron in the direction of $ {\bf {\cal B}}$ remain pinned to
this varying direction. However, owing to the rapid change in $\bf
v$ at the point of entry, the initial directions of $ {\bf {\cal
B}}$ are different for the two states that go around the
semi-circular rings from this point and interfere at the point of
exit. Hence, only the non-cyclic Berry phase corresponding to a
spin state going around a semi-circle, which is then completed to
a closed curve by the longitude, may be defined for each semi-
circular ring. Berry's phase is half the solid angle subtended by
the curve traced by $ {\bf {\cal B}}$ on the sphere that
represents all possible directions of the $ {\bf {\cal B}}$ and
completed into a closed curve by the geodesic joining the initial
and final directions at the center of this sphere. For unpolarised
electrons, as in the present experiment, the average of these
Berry phase factors may be observed.
\bigskip

\noindent
{\bf Acknowledgments}

\bigskip

I thank Stephen L. Adler for discussions and support at the Institute for Advanced Study, Princeton, NJ 08540, where part of this work was done. This work was also supported by a Fulbright Distinguished Scholar award, an ONR and NSF grants.


\begin{thebibliography}{99}

\bibitem{ah1959}
Y. Aharonov and D. Bohm, Phys. Rev. {\bf 115,} 485 (1959).

\bibitem{sa1994}
See, for example, J.J. Sakurai {\it Modern Quantum Mechanics}
(Addison- Wesley, 1994), p. 304.

\bibitem{zh2000}
S-L. Zhu and Z.D. Wang, Phys. Rev. Lett. {\bf 85,} 1076-1079
(2000).

\bibitem{ya2002}
Jeng-Bang Yau, E.P. De Poortere, and M. Shayegan, Phys. Rev. Lett.
{\bf 88,} 146801 (2002).

\bibitem{mo1998}
Analogous experiments were also performed by A.F. Morpurgo et al., Phys. Rev. Lett {\bf 80,} 1050 (1998); J. Nitta, H. Takayanagi, and S. Calvet, Microelectron. Eng. {\bf 47,} 85 (1999).

\bibitem{pe1989}
For a review of numerous theoretical analyses and experiments that
have resulted from the Aharonov-Bohm effect, see M. Peshkin and A.
Tonomura, {\it The Aharonov-Bohm Effect} (Springer-Verlag, 1989).

\bibitem{al1981}
B.L. Al'tshuler, , A.G. Aronov, and B.Z. Spivak, JETP Lett. {\bf
33,} 94-97 (1981); M. Buttiker, Y. Imry and R. Landauer, Phys.
Lett. A {\bf 96,}365 (1983).

\bibitem{wa1986}
See the review by S.Washburn and R. Webb, Adv. Phys. {\bf 35,}
375-422 (1986).

\bibitem{by1984}
Yu. A. Bychkov and E.I. Rashba, J. Phys. C: Solid State Phys. 17,
6039-6045 (1984).

\bibitem{be1984}
M.V. Berry, Proc. R. Soc. London A {\bf 392,} 45 (1984).

\end{thebibliography}
\end{document}